\documentclass[12pt,english,floatfix,superscriptaddress,aps,prd,preprint]{revtex4}
\usepackage[utf8]{inputenc}
\usepackage{amsmath}
\usepackage{amssymb}
\usepackage{amsbsy}
\usepackage{amsfonts}
\usepackage{amsopn}
\usepackage{amstext}
\usepackage{graphicx}
\usepackage{amssymb}
\usepackage{amsfonts}
\usepackage{amsmath}
\usepackage{amsmath,amsthm,amsfonts,amssymb}
\usepackage[mathcal]{eucal}
\usepackage{mathrsfs}
\usepackage{graphicx}
\usepackage[english]{babel}
\usepackage{color}
\usepackage{esint}
\usepackage[dvips]{epsfig}
\usepackage[dvips]{graphicx}
\usepackage{float}
\usepackage{units}
\usepackage{textcomp}

\usepackage{hyperref}
\usepackage{slashed}

\newcommand{\ie}{\begin{equation}}
\newcommand{\fe}{\end{equation}}
\newcommand{\se}{\begin{eqnarray}}
\newcommand{\ff}{\end{eqnarray}}

\begin{document}

\title{Reply on `Comment on The relativistic
Aharonov-Bohm-Coulomb system with
position-dependent mass'}


\author{R. R. S. Oliveira}
\email{rubensrso@fisica.ufc.br}
\affiliation{Universidade Federal do Cear\'a (UFC), Departamento de F\'isica,\\ Campus do Pici, Fortaleza - CE, C.P. 6030, 60455-760 - Brazil.}


\author{A. A. Ara\'ujo Filho}
\email{dilto@fisica.ufc.br}
\affiliation{Universidade Federal do Cear\'a (UFC), Departamento de F\'isica,\\ Campus do Pici, Fortaleza - CE, C.P. 6030, 60455-760 - Brazil.}


\author{R. V. Maluf}
\email{r.v.maluf@fisica.ufc.br}
\affiliation{Universidade Federal do Cear\'a (UFC), Departamento de F\'isica,\\ Campus do Pici, Fortaleza - CE, C.P. 6030, 60455-760 - Brazil.}


\author{C. A. S. Almeida}
\email{carlos@fisica.ufc.br}
\affiliation{Universidade Federal do Cear\'a (UFC), Departamento de F\'isica,\\ Campus do Pici, Fortaleza - CE, C.P. 6030, 60455-760 - Brazil.}

\date{\today}

\begin{abstract} It is shown that the results of our paper are correct, although the use of the chirality operator $\gamma_5=\sigma_1$ in the paper is not correct, as fairly the authors of the Comment stated.
\end{abstract}

\maketitle


In the \textit{Comment on The relativistic
Aharonov-Bohm-Coulomb system with
position-dependent mass},  Mendrot and Castro \cite{Mendrot} have mentioned one criticism and one suggestion about our paper, and they claimed that the main conclusion of our article was impaired.

The criticism is the following:

\textit{ The fundamental property $P_R\gamma^\mu=\gamma^\mu P_L$ is not truthful in the $(2+1)$-dimensional spacetime. Therefore, the chirality operator $\gamma_5=\sigma_1$ does not anticommute with all the $\gamma^\mu$-matrices appearing in the Dirac equation. Therefore, the fulcrum of that paper are the $left-handed$ $P_L=(I_{2\times 2}-\gamma^5)/2$ and $right-handed$ $P_R=(I_{2\times 2}+\gamma^5)/2$ projection operators.}
 
 The suggestion is the following:

\textit{However, the physical implications of the studied system might be considered by replacing the original first-order Dirac equation with a pair of coupled first-order differential equations for the upper and lower components of the Dirac spinor, or by multiplying both sides of Eq. (1) by $\gamma^\mu\Pi_\mu+m({\bf r})c$ in order to find a second-order differential equation.} 

Our response to the criticism is:

In $(2+1)$-dimensional spacetime, the $\gamma^\mu$ gamma matrices that appear in the Dirac equation are the $2\times 2$ Pauli matrices and the chirality matrix $\gamma_5$ is simply the Pauli matrix $\sigma_1$. Indeed, the property $P_R \gamma^\mu= \gamma^\mu P_L$ is not entirely satisfied for the case $(2+1)$-dimensional (is only partially satisfied), where $\gamma_5$ anticommutes with $\sigma_2$ and $\sigma_3$, but not anticommutes with $\sigma_1$. Therefore, in this point, the criticism is completely right.

However, the physical implications of the studied system might be considered ( the energy spectra do not changes) if we define our original Dirac spinor in terms of another one. Explicitly, we have \cite{Kluger,Vakarchuk}
\ie \Psi({t,\bf r})\equiv[\gamma^\mu\Pi_\mu+m({\bf r})c]\psi({t,\bf r}), 
\label{1}\fe

In particular, this Dirac spinor is used to study physical problems involving fermion pair production in strong electric fields \cite{Kluger} and the Kepler problem for a particle with position-dependent mass \cite{Vakarchuk}. A justification for the definition of this spinor comes from the fact that the (free) quadratic Dirac equation is the Klein-Gordon equation \cite{Bentez}, and one of the consistent ways to achieve this is through the definition \eqref{1}. For example, as the free Dirac equation is given by $(\gamma^\mu p_\mu-m_0 c)\Psi=0$, and defining $\Psi=(\gamma^\mu p_\mu+m_0 c)\psi$, results in $(\gamma^\mu p_\mu-m_0 c)(\gamma^\nu p_\nu+m_0 c)\psi=(p^\mu p_\mu-m^2_0c^4)\psi=0$, where we use the properties $\{\gamma^\mu,\gamma^\nu\}=2g^{\mu\nu}$ and $p^\mu=g^{\mu\nu}p_\nu$. In addition, it is important to note that definition \eqref{1} is general, i.e., it applies to any dimension of spacetime \cite{Kluger,Vakarchuk}.

Therefore, using the definition of spinor in Eq. \eqref{1} results in a quadratic Dirac equation exactly equal to our paper. At this point, the energy spectra are not modified (more relevant results). On the other hand, the Dirac spinor given by Eq. (33) in our paper must not contain the term $1/m(\rho)c$ anymore. So, from Eq. \eqref{1} here we must do the following changes in our paper to avoid confusions: $\Psi_R({t,\bf r})\to\psi({t,\bf r})$, and $\psi_R(t,\rho,\theta)\to\chi(t,\rho,\theta)$, with $\chi(t,\rho,\theta)\equiv U^{-1}(\theta)\psi(t,\rho,\theta)$.

\section*{Acknowledgments}
The authors thank the fair comment.

\end{document}